\documentclass[preprint2,usenames,dvipsnames]{aastex6}

\usepackage[free-standing-units]{siunitx} 
\usepackage{natbib} 

\usepackage{multirow} 
\usepackage{paralist}
\usepackage{xspace} 
\usepackage{rotating} 
 \usepackage{epsfig}
 \usepackage{subfigure}
 \usepackage{epstopdf}
\usepackage{graphicx}
\usepackage{amsmath}
\newcommand{\av }{AV1224\xspace}

\shorttitle{Publications of the Astronomical Society of the Pacific \hspace{5cm}  Fox-Machado et al. }
\shortauthors{Fox-Machado et al.}

\begin{document}

\title{Photometric and spectroscopic observations of the neglected near-contact binary Cl*~Melotte~111~AV~1224\footnote{Based upon
    observations acquired at: Observatorio Astron\'omico Nacional in
    the Sierra San Pedro M\'artir (OAN-SPM), Baja California,
    M\'exico and Xing Long Station in China.}}




\author{L.~Fox-Machado\altaffilmark{1},
  T.Q.~Cang\altaffilmark{2},
  R.~Michel\altaffilmark{1}, 
  J.N.~Fu\altaffilmark{2}, 
  C.Q.~Li\altaffilmark{2}, 
   }

\altaffiltext{1}{
    Observatorio Astron\'omico Nacional, Instituto de Astronom\'ia --
  Universidad Nacional Aut\'onoma de M\'exico, Ap. P. 877, Ensenada, BC
  22860, Mexico
\email{lfox@astro.unam.mx}
}

\altaffiltext{2}{
Department of Astronomy, Beijing Normal University, No. 19 Xinjiekouwai Street, 
Haidian District, Beijing 100875, China
    }

\begin{abstract}

This paper presents a photometric and spectroscopic study of the short-period binary star 
 Cl*~Melotte~111~AV~1224. Measurements in the $B$, $V$, and $R$ passbands obtained during three observing runs 
 between 2014 and 2017 and medium-resolution spectra secured in 2014, are analyzed together with public data from
 the SuperWASP and LAMOST projects.  Our light curves show marked asymmetry with a variable O'Connell effect.  The SuperWASP photometry 
 is used to derive a mean binary period of 0.345225 days.
 The analysis of the $(O-C)$ diagram reveals that the orbital period is decreasing at a rate of $dP/dt = -3.87 \times 10^{-6}$ days yr$^{-1}$, which
 may be caused by mass transfer from the more-massive component to the less-massive one.
 The system is found to be a single-lined spectroscopic binary with
a systemic velocity, $\gamma =  1 \pm 3$ Km s$^{-1}$,  and a semi-amplitude, K$_{1}$  = 21 $\pm$ 5 Km s$^{-1}$. 
The spectral classification and the effective temperature of the primary component are estimated to be  K0V $\pm$ 1 and $5200 \pm 150$ K, respectively.
The photometric and spectroscopic solutions reveal that Cl*~Melotte~111~AV~1224 is a low-mass ratio ($q=m_{2}/m_{1} \sim 0.11$), 
low-inclination  ($\sim ~ 38^{\circ}$)  near-contact system. The masses, radii and luminosity for the primary and secondary
are estimated to be $1.02 \pm 0.06\, M_\odot$, $1.23 \pm 0.05\, R_\odot $,  $1.01 \pm 0.06\, L_\odot$ and $0.11  \pm 0.08\, M_\odot$, $0.45 \pm 0.05\, R_\odot$, $0.10 \pm 0.06\, L_\odot$,
respectively.  The marginal contact, together with the period decrease, suggests that this binary system may be at a key evolutionary stage,
as predicted by the theory of thermal relaxation oscillations.

\end{abstract}


\section{Introduction}
\label{sect:1}

Near-contact binaries (NCBs) are a important source in understanding the formation
and evolution of the binary systems, as they are assumed to be lying in a key evolutionary stage. 
They have been defined by \citet{shaw} as a subclass of close binaries 
in which both components fill, or nearly fill, their critical Roche lobe and 
are near enough to each other as to have strong proximity effects like those of W~UMa type systems without being in contact.

NCBs have periods of less than a day, display strong tidal interaction, and show EB-type variations. Such systems
may be the evolutionary precursors to the A-type W~UMa systems, and are probably in the early stages
of mass transfer. Indeed, some NCBs exhibit asymmetric light curves and orbital period decrease 
caused probably by mass transfer or/and by angular momentum loss via magnetic braking. As the period decreases, 
the separation between both components will be reduced, and the system will evolve into an A-type overcontact binary. 
 Therefore,  it is suggested that the NCBs could be the missing link between
detached binaries and W~UMa systems.   There is a considerable number of newly discovered 
  near-contact systems brighter than 12 mag $(V)$ since the early work by \cite{shaw}, which are divided into two subclasses:  V1010 Ophiuchi systems and
 FO Virginis binaries.  The study of fainter NCBs ($V > 13$ mag) is challenging, but also important, because they may provide new insights into 
underlying mechanisms of the mass transfer in close binaries.

\begin{table*}
\caption{Log of Photometric and Spectroscopic Observations.}
\begin{tabular}{|lccccr|ccc|}
\hline
\hline
\noalign{\smallskip}
\multispan6 \hfill Photometric Observations \hfill \vline & \multispan3 \hfill Spectroscopic Observations \hfill  \\
 \noalign{\smallskip}
\hline
    UT Date  & Start Time  & End Time &  Filter       &OAN-SPM &XL     & Start Time &Number of spectra & Exp. Time  \\
              &(HJD 2454900+)&(HJD 2454900+)&         &  (hr) & (hr)   &  (HJD 2454900+)              &   &  (sec)              \\
 \hline
       2014 April 05   & 1853.00  &   1853.34 &  $BVR$  & -     & 8.1 &   -    &     -        &         -          \\
       2014 April 06   & 1854.00  &   1854.34 &  $BVR$  & -     & 8.3 &   -    &   -          &      -            \\
       2014 April 08   & 1855.63  &   1856.01 &  $BVR$  & 9.2   & -   & 1855.70 & 6                  & 1800                \\
       2014 April 09   & 1856.63  &   1857.25 &  $BVR$  & 7.6   & 7.6  & 1856.70&  8                &  1800                \\
       2014 April 10   & 1857.90  &   1857.99 &  $BVR$  & 2.3   & -   &  1857.90&  4                &   1800               \\
       2016 April 15  & 2593.75  &    2594.00 &  $BVR$  & 7.9   &-    &   -      &      -    &    -               \\
       2016 April 17  & 2595.64  &    2595.86&  $BVR$  & 5.3   &-   &    -       &   -       &   -              \\   
       2016 April 18  & 2596.67  &    5597.00&  $BVR$  & 7.6   &-  &    -        &   -       &   -                \\
       2016 April 20  & 2598.68  &    2598.98&  $BVR$  & 7.3   &-  &    -        &    -      &     -          \\ 
       2017 March 28  & 2940.64  &    2940.93 &  $BVR$ & 6.9   & - &     -       &    -     &     -          \\
       2017 April 13  & 2956.64  &     2967.00& $BVR$&  8.7   & - &     -        &    -    &      -          \\  
 \hline
\end{tabular}
\label{tab:log}
\end{table*}

In this paper, we present a detailed analysis of Cl*~Melotte~111~AV~1224\footnote{Star designation in the Simbad astronomical database. AV~1224 refers to the star running number 
in the astrometric catalog by \cite{abad}.}  (GSC 01989-00964, 2MASS~J12214222+2500569, and \av hereater ) --a NCB candidate 
 target ($V \sim 13.4$ mag) identified as a short-period eclipsing binary by \citet{fox}. 
The light curves in the $V$ band show $\beta$ Lyrae type variations  and a remarkable O'Connell effect. 
Because the preliminary photometric solutions reported by \citet{fox} were unable to discriminate
between overcontact and semi-detached configurations, we have conducted follow-up spectroscopic and photometric 
optical observations of \av to shed more light on its nature. We incorporated  publicly available SuperWASP and LAMOST data into our study. 
The absolute physical parameters of the components are derived, and the evolutionary status of this rare binary system is discussed.


\section{Observations}
\label{sect:2}

Table~\ref{tab:log} provides relevant information about our photometric and spectroscopic observations. 
 CCD photometric data were collected at the Observatorio Astron\'omico Nacional at San Pedro M\'artir (OAN-SPM hereafter), 
in Baja California, Mexico. In 2014, additional photometry was obtained at 
Xinglong Station of the National Astronomical Observatories of China (XL hereafter).
 At OAN-SPM observatory, we used the same procedures as described in \citet{fox}. 
Briefly, the CCD  images were acquired with the 0.84 m f/15 Ritchey-Chr\'etien telescope. 
The telescope hosted the `Mexman' filter wheel and a Marconi (E2V) CCD camera,
which has a 2048 $\times$ 2048 pixels array, with a pixel size of 13.5 $\times$ 13.5 $\mu$m$^{2}$.
The gain and readout noise of this CCD camera are 1.8 e$^{-}$/ADU and 4.7 e$^{-}$,
respectively.   The field of view in this configuration  
is about 8$^{\prime}$ $\times$ 8$^{\prime}$ arcmin$^{2}$  with the
scale of 0$^{\prime \prime}$.22/pixel with binning factor $1 \times 1$.
At XL,  we used the 0.85-m telescope with a PI $1024 \times 1024$ TE CCD camera mounted on the primary focus
of the telescope. The effective field of view is $8^{\prime}.5 \times 8^{\prime}.5$  arcmin$^{2}$. In all seasons, the observations were made 
in the $B$, $V$, and $R$-bands  with exposure times normally set 
at 80 s, 40 s, and 30 s, respectively, depending on the weather conditions.  
Sky flats, dark and bias exposures were also taken regularly. 

\begin{figure}[]
\centering
\includegraphics[width=8cm]{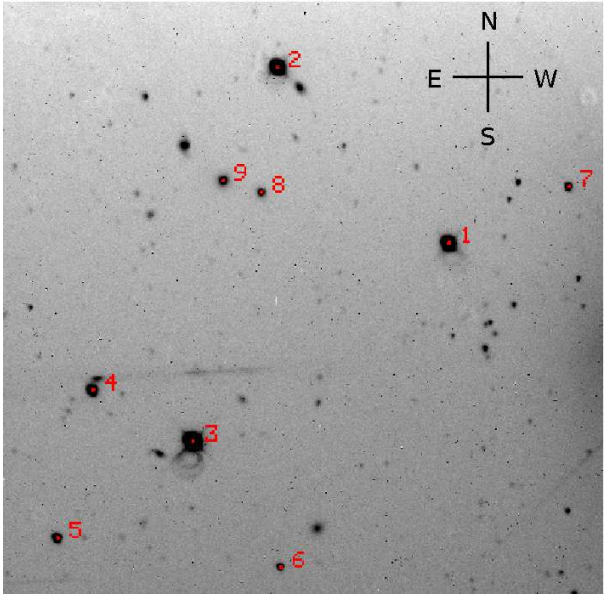}
\caption{Image of the CCD field of view ($8' \times 8'$) at the 0.84 m telescope at OAN-SPM observatory. 
The standard photometry of the stars is given in Table~\ref{tab:par}. North is up and east is left.}
\label{fig:field}
\end{figure}

The images were reduced in the standard way, using the IRAF\footnote{Image Reduction and Analysis Facility} package. 
The reduction process included  overscan correction,  bias subtraction, and  flat field correction. 
The cleaning of cosmic rays was made with the external task {\it lacos} \citep{vandokkum}.  The 
instrumental magnitudes of the stars  were computed using the aperture photometry method implemented in 
the IRAF/DAOPHOT package. Figure~\ref{fig:field} shows a typical image taken with the 0.84 m telescope of 
the OAN-SPM observatory.  For stars labeled with digits 1 through 9 the Table~\ref{tab:par} 
provides the star names, positions, and our own $UBVRI$  standard photometry measurements. 
To determine the relative magnitudes (and colors) of the
stars, the field was measured during a photometric night
along with some Landolt standards at different airmasses. Transformation
equations were calculated and the resulting magnitudes are shown in
Table~\ref{tab:par}.
Differential photometry was made with respect to Cl*~Melotte~111~AV~1236  which is comparable in brightness and color to the target star. 
Cl*~Melotte~111~AV~1248 was used as check star. They are labeled with numbers 2 and 3 in Figure~\ref{fig:field} respectively. 
 No evidence of periodic variation among the comparison and check stars was found. The light curves were inspected, and
 those high-dispersed data points due to clouds and bad weather conditions were removed.
  A total of 18 spectra of \av 
were acquired over three nights in 2014 on the 2.12 m telescope at
OAN-SPM observatory. The spectra were obtained with the 
Boller \& Chivens spectrograph  installed in the Cas\-se\-grain focus of the telescope.
A 1200 grooves mm$^{-1}$ grating  and a 13.5 $\mu$m 2048$\times$2048 pixel E2V Marconi CCD camera 
to cover a wavelength range from 4400\,{\AA} to 6700\,{\AA}, were used in the observations.
The typical spectral resolution of the recorded spectra is 1.8\,{\AA}. The data reduction was made with the standard long-slit routines
   of the IRAF package after applying bias subtraction. The cleaning of cosmic rays was made with the
   external task {\it lacos} \citep{vandokkum}. The wavelength calibration was made using a {\rm CuHeNeAr} arc lamp 
    before and after every object spectra.  The spectra  were flux calibrated using spectrophotometric standard stars 
    observed in the same night.

\begin{figure}[!t]
\includegraphics[width=9.5cm]{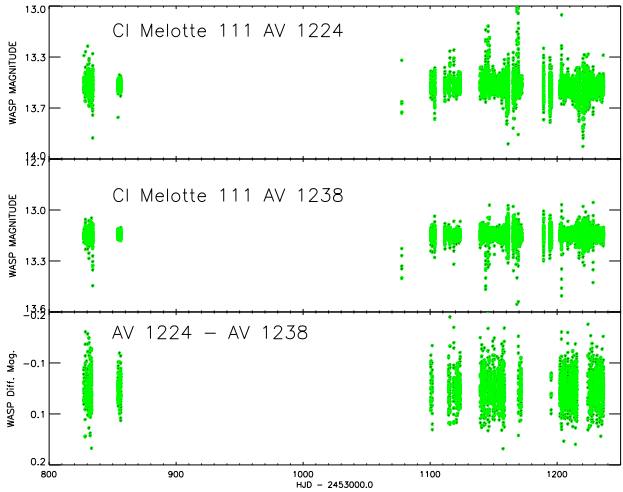}
\caption{SuperWASP light curves of \av (top panel), Cl*~Melotte~111~AV~1236 (middle panel) and the differential magnitudes (low panel).}
\label{fig:swlc}
\end{figure}

The SuperWASP light curves of \av and  Cl*~Melottte~111~AV~1238 stars are displayed 
in the upper and middle panels of Figure~\ref{fig:swlc}, respectively.  They comprise 6300 measurements taken
between  2004 May and 2008 May. The stars are rather faint and, due to
the small aperture telescopes employed by the SuperWASP project,
they exhibit relatively large scatter\footnote{SuperWASP photometry achieves a 
precision of better than 1\%  for objects with magnitudes $7 \le V \le 11.5$}. For this reason, the star was neglected for further study.  
Although our data have better precision,
the SuperWASP timeseries covering a much longer time span are well suited for period determination purposes.
 To reduce light-curve scatter, we excluded the data points with errors greater than 0.08 mag as given by SuperWASP project.  
       Then, we computed the difference in magnitudes between \av with 
   respect to Cl*~Melotte~111~AV~1236.  The resulting differential light curves, which contain $\sim$ 2950 measurements (see lower panel of Fig.~\ref{fig:swlc}), 
 are used to estimate the period of the binary star.

\begin{center}
\begin{table*}
\caption{Calibrated $UBVRI$ Photometry for Stars in the \av Field. }
\begin{tabular}{l c c c c c c c}
\hline
\hline
  Star ID & RA  & DEC &  $U$ & $B$ & $V$ &$ R$ & $I$ \\
          &J2000.0&J2000.0&(mag)&(mag)&(mag)&(mag)&(mag)   \\
 \hline
 1-\av& \footnotesize 185.425923 &\footnotesize $+$25.015790&\footnotesize $14.443 \pm 0.004$ & \footnotesize $14.153 \pm  0.001  $   &  \footnotesize $13.413 \pm  0.002  $     & \footnotesize $ 12.967 \pm 0.002 $ &\footnotesize $12.584 \pm 0.001$ \\
  
 2-{\footnotesize Cl*~Melotte~111~AV~1236}& \footnotesize 185.467183 & \footnotesize $+$25.054349 & \footnotesize $14.187 \pm 0.003$& \footnotesize $13.754 \pm 0.002 $    & \footnotesize $13.020 \pm 0.003  $  &\footnotesize $12.610  \pm 0.004 $&\footnotesize $ 12.279 \pm 0.004$\\
 
  3-{\footnotesize Cl*~Melotte~111~AV~1248} &\footnotesize 185.487535& \footnotesize $+$24.972642& \footnotesize $13.654 \pm 0.002$ & \footnotesize $13.201 \pm   0.004$&\footnotesize  $12.394  \pm 0.003  $ &\footnotesize  $11.925 \pm 0.003 $& \footnotesize$11.539 \pm 0.002$\\
  
  4-{\footnotesize 2MASSJ12220283$+$2459016}&\footnotesize 185.511714& \footnotesize $+$24.983695&\footnotesize $ 16.602 \pm 0.001$ &\footnotesize  $16.059 \pm 0.003 $&  \footnotesize $15.202  \pm 0.001  $ &\footnotesize $14.688 \pm 0.002  $ & \footnotesize $ 14.255 \pm 0.002$\\  
  
   5-{\footnotesize 2MASSJ12220484$+$2457053}& \footnotesize  185.520269 &\footnotesize $+$24.951456 & \footnotesize $ 17.392 \pm 0.006$ & \footnotesize $17.056 \pm 0.001 $& \footnotesize $16.292  \pm 0.002  $& \footnotesize $15.786 \pm 0.003 $ & \footnotesize $15.403 \pm 0.002$\\

  6-{\footnotesize 2MASSJ12215192$+$2456417} &\footnotesize 185.466526&\footnotesize $+$24.945096& \footnotesize $18.042 \pm 0.003$& \footnotesize $ 18.195 \pm 0.001 $& \footnotesize $17.683 \pm  0.003 $ & \footnotesize$17.314 \pm 0.002 $& \footnotesize $ 17.002 \pm 0.004$\\

  7-{\footnotesize 2MASSJ12213524$+$2501411}& \footnotesize 185.396903& \footnotesize$+$25.028115 & \footnotesize $17.808 \pm 0.002$& \footnotesize $17.794 \pm  0.002$& \footnotesize $17.122 \pm 0.002  $& \footnotesize$ 16.686  \pm 0.001$& \footnotesize $16.267  \pm 0.002$\\

 8-{\footnotesize 2MASSJ12215307$+$2501365}  &\footnotesize 185.471147& \footnotesize$+$25.026827 &  ... & \footnotesize $19.655 \pm 0.001 $& \footnotesize  $18.319 \pm 0.007$ & \footnotesize $17.652  \pm 0.005 $&\footnotesize $17.062 \pm 0.006$\\

9-{\footnotesize 2MASSJ12215525$+$2501460} &\footnotesize 185.480268& \footnotesize $+$25.029474 & \footnotesize $19.763 \pm 0.008$  &  \footnotesize$ 19.296 \pm 0.024 $&  \footnotesize $17.930 \pm 0.011 $ &\footnotesize $17.325 \pm 0.009 $& \footnotesize $16.718 \pm 0.017$\\ 
 \hline
\end{tabular}
\label{tab:par}
\end{table*}
\end{center}


\section{Data analysis}   
\label{sect:3}

\subsection{Spectroscopic Analysis}
\label{sec:spec}

The radial velocities were measured with the IRAF task {\rm fxcor}, by cross-correlating
 the spectra against the spectra of radial-velocity standard stars  HD~38230 \citep{udry} and HD~132734 \citep{stefanik}.  
 The whole spectral ranges were used in the cross-correlation process. 
 The weighted mean values of radial velocities, with the uncertainties 
 obtained with {\rm fxcor}, are given in Table~\ref{tab:rv}.  No signs of the secondary
 spectrum have been detected. The radial velocity curve, which is related to the  more-massive component, 
 folded with the photometric period is shown in Figure~\ref{fig:ph2} (the bottom panel of the first column). 
 The radial-velocity light curve shows that the massive component is moving toward us and the less-massive one, not seen, is moving away from
us after the primary eclipse at phase 0.0. Therefore, the more-massive star is eclipsed at primary minimum and it is the hotter component.
   We used the  {\rm RVFIT}\footnote{http://www.cefca.es/people/$\sim$riglesias/rvfit.html} IDL code \citep{iglesias} to fit a Keplerian
  orbit to the data. The best fit is shown by continuous lines in the bottom panel of first column of Fig.~\ref{fig:ph2}.  We 
  obtained  $\gamma$ = 1 $\pm$ 3  Km s$^{-1}$ and K$_{1}$ = 21 $\pm$ 5  Km s$^{-1}$. 
  The LAMOST database lists for \av a radial velocity of 8.42 Km s$^{-2}$ taken at HJD 2455966.31812. 
     This single radial-velocity value, which is shown with a rhombus in Figure~\ref{fig:ph2}, is
     in good agreement with our measurements.   
      
 We have derived a spectral classification by comparing a phase 0.5 program spectrum with
those spectra of MILES library \citep{sanchez}, suitably smoothed to our resolution, following the procedure explained
by \citet{baran1}. We assigned a K0V $\pm$ 1 spectral type.
On the other hand, the 2MASS photometry of \av gives $J = 11.937$, $H = 11.589$ and $K_{s} = 11.517$ \citep{skrutskie}.
 Using the simplified equations of \citet{koen},  the 2MASS indices $(J-H) = 0.348$ and $(H-K_{S}) = 0.072$  can be transformed 
 into $(J-H) = 0.421$, $(H-K) = 0.051$ in the SAAO system. 
 According to the calibrations of spectral types on SAAO near-infrared indices by \citet{bessell},
these infrared photometric indices imply spectral types G6V-K0V.

A grid of synthetic spectra was computed with the {\it iSpec}\footnote{https://www.blancocuaresma.com/s/iSpec} code \citep{blanco} and 
then cross-correlated  to a phase 0.5 target spectrum. The best matching template had a temperature of 5200 $\pm$ 150 K and $\log\;g=3.7 \pm 0.5$ cm s$^{-2}$.
These values are in agreement with the parameters of \av listed by the LAMOST team.
The LAMOST spectra are automatically analyzed by instrument-specific software to determine fundamental parameters of the star. 
 For \av, the pipeline suggests $T_{\rm eff} = 5166$ K, $\log\, g = 3.7 $ cm s$^{-2}$ .

\begin{figure}[!t]
\includegraphics[width=8.5cm]{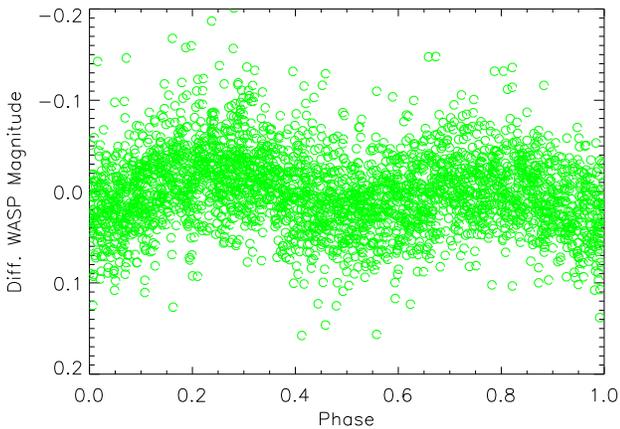}
\caption{SuperWASP differential light curves of \av folded at $P=0.345225$.}
\label{fig:pha0}
\end{figure}

\subsection{Period Analysis and Determination of Ephemeris}
\label{sec:per}

\begin{figure}[!b]
\centering
\includegraphics[width=12.5cm]{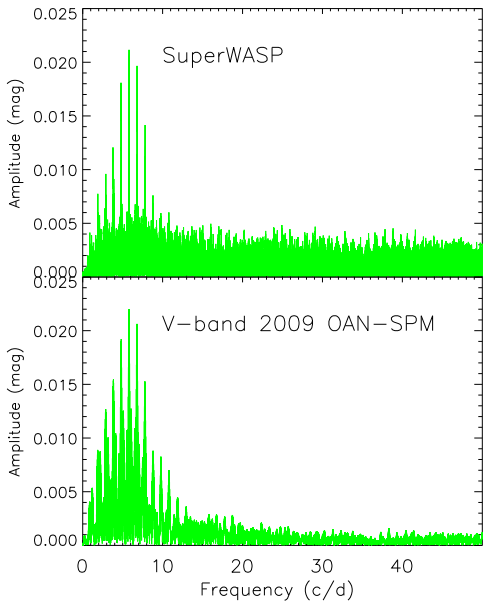}
\caption{Top: amplitude spectrum from the SuperWASP data. Bottom: amplitude spectrum from the 2009 OAN-SPM  $V-$band data. }
\label{fig:spec}
\end{figure}

The {\rm Period04} package \citep{lenz} was employed to compute the period of the binary system based on the SuperWASP differential timeseries
  obtained according to the steps described in the previous section. The eclipsing binary light curves can be represented
by two sine waves with frequencies $\nu_{1} \sim 5.8$ c/d and $\nu_{2} \sim 2.9$ c/d.   
Multiplying the period by 2 gives the appropriate orbital period of the binary system as $P=0.345225 \pm 0.000005$.
The periodogram of the SuperWASP timeseries is shown in the upper panel of Figure~\ref{fig:spec}. 
As a comparison,  the periodogram of the 2009 $V$-band observations reported by \citet{fox} is depicted in the lower panel.
The SuperWASP data folded at period $P=0.345225$  are displayed in Figure~\ref{fig:pha0}.

Seven eclipses of \av, including four primary eclipses
and three secondary ones, were recorded during our observations.
The epochs of these light minima were determined via the method described by \citet{kwee} and the
 mean values of $B$, $V$ and $R$ filters are listed in Table~\ref{tab:min}. Using these data and the times 
 of minima reported by \citet{fox}, we derived the following
revised linear and quadratic ephemeris for the binary system
by applying the least-squares method:

\begin{equation}
HJD_{\rm min}\,I =  2454941.7528(6) +   0.34522(5) \times E
\end{equation}

\begin{equation}
\begin{split}
 HJD_{\rm min}\,I =  2454941.7435(9) +   0.34522(4) \times E  \\
 - 1.8(3)\times 10^{-9} \times E^{2}
\end{split}
 \end{equation}

\begin{figure}[]
\includegraphics[width=8.5cm]{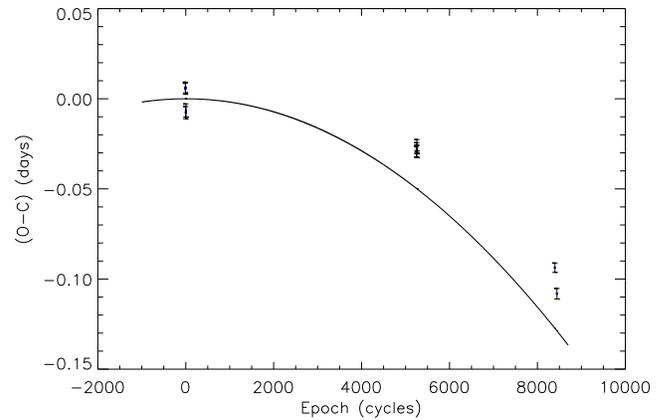}
\caption{$(O-C)$ curve for the eclipsing binary \av. The solid line represents the parabolic fit from equation (2).}
\label{fig:oc}
\end{figure}

 \noindent Equation (2) implies that the orbital period of the \av decreases secularly at a rate of
 $dP/dt = -3.87 \times 10^{-6}$ days yr$^{-1}$. However, due to the small number of minimum times, a cyclic period
 variation cannot be ruled out.  { The $(O-C)$ diagram is plotted in Figure~\ref{fig:oc}.}
 { The SuperWASP data were not used in the O-C analysis because, due to their much large scatter,  the times of minimum could not be determined.}

\begin{table}[!t]
\caption{Radial Velocities for \av. \label{tab:rv}}
\begin{tabular}{lcr}
 \hline
 \hline
HJD & Phase & RV \\
    &       & (Km s$^{-1}$) \\
    \hline
2456755.73313 &  0.31 &   -11.9    $\pm$   24.7  \\
2456755.76568 &  0.42 &    -4.3   $\pm$      21.3 \\
2456755.78889 &  0.49 &    1.3    $\pm$   16.6    \\
2456755.83411 &  0.63 &      6.5  $\pm$       13.1  \\
2456755.90133 &  0.84 &     23.2   $\pm$      9.6   \\
2456755.92735 &  0.92 &     10.7  $\pm$      10.3   \\
2456756.70119 &  0.34 &    -10.1   $\pm$    27.3   \\
2456756.72935 &  0.43 &    -1.2   $\pm$      17.4   \\
2456756.75229 &  0.50 &      7.7   $\pm$     12.8    \\
2456756.77439 &  0.56 &     15.2    $\pm$    14.3    \\
2456756.82153 &  0.71 &     14.7   $\pm$     11.4    \\
2456756.89992 &  0.96 &     12.7    $\pm$    13.9    \\
2456756.9228  &  0.03 &      -10.0   $\pm$     14.3   \\
2456756.97229 &  0.19 &      -12.0   $\pm$     12.2   \\
2456757.89086 &  0.06 &    -20.8   $\pm$    16.7     \\
2456757.91319 &  0.13 &     -3.2    $\pm$    18.7    \\
2456757.93563 &  0.20 &    -17.5    $\pm$   17.2    \\
2456757.97966 &  0.33 &    -29.12    $\pm$    10.6   \\  
 \hline
\end{tabular}
\end{table} 
  
\begin{table}[!t]\centering
  \setlength{\tabcolsep}{1.0\tabcolsep}
 \caption{New Times of Minima for \av.\label{tab:min}}
  \begin{tabular}{lcc}
\hline
\hline
HJD      &      Minimum Type   & Filter     \\
       
\hline
$2456753.15126 \pm 0.00225$  & I & $BVR$ \\
 $2456755.75641 \pm 0.00212$ & II & $BVR$ \\
$ 2456755.91444 \pm   0.00205$ & I&   $BVR$  \\
 $ 2456756.79359 \pm  0.00229$ & II  & $BVR$\\
 $2456756.95242 \pm  0.00315$ & I  &  $BVR$\\
$ 2457840.80406 \pm  0.00217 $  &II &  $BVR$\\
 $2457856.86172  \pm   0.00255$  &I &  $BVR$ \\

\hline
\end{tabular}
\label{tab:max}
\end{table}
 
\begin{figure}[!t]
\includegraphics[width=8.5cm]{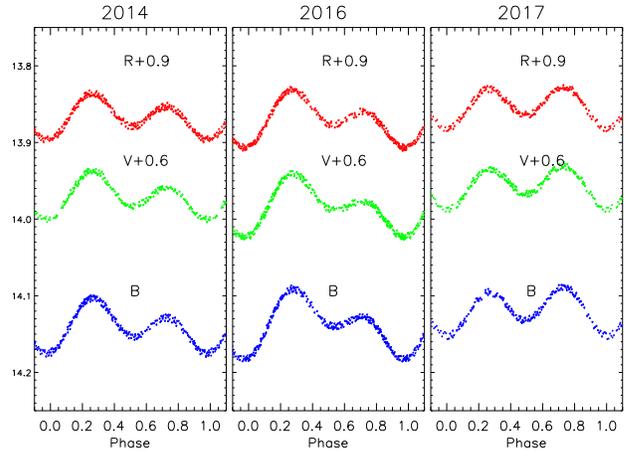}
\caption{$B$, $V$, and $R$  phase-folded light curves of \av with a period of 0.345225 days observed in 2014 (left panels), 2016 (central panels), and 2017 (right panels).}
\label{fig:ph1}
\end{figure}

\begin{table*}
\caption{ $BVR$-band Observations for Eclipsing Binary \av. The differential
Magnitudes can be Obtained by Subtracting the $BVR$ Magnitudes of Cl*~Melotte~111~AV~1236 Listed in Table~\ref{tab:par} from
the $BVR$ Relative Magnitudes Listed Here.}
\begin{tabular}{cccccc}
\hline
\hline
  HJD & $m_{B}$  & HJD  &  $m_{V}$ & HJD &  $m_{R}$  \\
      &(mag)&&(mag)&&(mag)   \\
 \hline
 2456752.99733 &14.1574  &2456752.99904   &13.3747  &2456753.00007  &12.9802 \\
 2456753.00163 &14.1558  &2456753.00765   &13.3777  &2456753.00437  &12.9736 \\
 2456753.01212 &14.1565  &2456753.01384   &13.3754  &2456753.00868  &12.9804 \\
 2456753.01785 &14.1570  &2456753.02009   &13.3822  &2456753.01486  &12.9673 \\
 2456753.02355 &14.1517  &2456753.02579   &13.3754  &2456753.02147  &12.9676 \\
 2456753.02925 &14.1484  &2456753.03149   &13.3718  &2456753.02716  &12.9647 \\
\hline
\multicolumn{6}{l}{ Note:Table~\ref{tab:mag} is published in its entirely in the electronic edition of the \pasp.}\\
\multicolumn{6}{l}{ A portion is shown here for guidance
regarding its form and content.} \\
\end{tabular}
\label{tab:mag}
\end{table*}

\begin{table*}
\caption{Characteristics Parameters of the Light Curves of \av. Primary and Secondary Maxima (Max I and Max II) are Defined
as that following the Primary and Secondary Minima, Respectively.}
\begin{tabular}{lccc|ccc|ccc}

\hline
\hline

          &  2014$B$ &   2014$V$ &     2014$R$  & 2016$B$  &    2016$V$  &   2016$R$ & 2017$B$ & 2017$V$ &  2017$R$\\  
\hline          
          
Min I\; (mag)   &  14.175    &   13.400  &    12.996   &  14.182 &   13.424    & 13.007  &  14.153 & 13.389  &  12.982\\

Min II (mag)    & 14.153   &   13.383   &    12.978  &  14.139  &  13.387   &  12.975   &  14.132 & 13.368   & 12.965\\

Max I  (mag)   &  14.103 &     13.338   &   12.936 &  14.092  &    13.341 &    12.932  &  14.096  & 13.335 &12.930\\
                               
Max II  (mag) &   14.128  &    13.359  &    12.954 &  14.128   &   13.377  &  12.961  &  14.089  &  13.331 &12.928\\ 
\hline
Min I - Min II  (mag) &    $+$0.022  &    $+$0.017 &  $+$0.018  & $+$0.043 &  $+$0.037  &  $+$0.032 &$+$0.021 &$+$0.021 & $+$0.017 \\

Max II - Max I  (mag)&     $+$0.025 &    $+$0.021 & $+$0.018  &  $+$0.036 &   $+$0.036  & $+$0.029   & $-$0.007& $-$0.004& $-$0.002\\

Min I - Max I\; (mag)&$+$0.072 &  $+$0.062 & $+$0.060 & $+$0.090& $+$0.080 &$+$0.075& $+$0.057 & $+$0.054& $+$0.052\\

Min II - Max II (mag)& $+$0.025& $+$0.024 & $+$0.024 & $+$0.011&  $+$0.010 & $+$0.014 & $+$0.046& $+$0.037 & $+$0.037\\

\hline
\hline
\end{tabular}
\label{tab:min_max}
\end{table*}

\subsection{The Asymmetry and Variations of the Light Curves}
\label{sec:var}
The phased $B$, $V$, and $R$ light curves folded using Equation (1) for  \av are shown in Figure~\ref{fig:ph1}.
 The light curves are plotted  in terms of their relative magnitudes based on the reference star Cl*~Melotte~111~AV~1236.
 A total of 2674 individual observations in three bands (i.e.,
897 in $B$, 895 in $V$ , and 882 in $R$) are tabulated in Table~\ref{tab:mag} in the
form of HJD time versus magnitude.

  It can be seen that the maxima, the minima, and the general shape of the light curves changed over seasons. In particular,
  dramatic changes occur around Max II (at 0.75 phase). In Table~\ref{tab:min_max} we list, for each color band, 
  the magnitudes of the maxima and minima; the depths of the primary and the secondary eclipses  (Min I - Max I and Min II - Max II); 
  the magnitude difference between depths of primary and secondary minima
(Min I - Min II); and the magnitude of the difference between the amplitude of the secondary and the primary maxima (Max II - Max I). 
From Table~\ref{tab:min_max} and from Figure~\ref{fig:ph1}, it follows that in 2014 the light curves showed a positive O'Connell effect with 
the Max I brighter than the Max II 
by 0.025, 0.021, 0.018 mag in the $B$, $V$ and $R$ bands respectively. Two years later in 2016, the O'Connell effect was still positive  
but as the Max I in 2016 becomes brighter than Max I in 2014 and  Max II in 2016 becomes fainter than the Max II in 2014,
the amplitude of the Max II - Max I in 2016 grew higher by 0.036, 0.036 and 0.029 in 
the  $B$, $V$, and $R$ bands respectively. In 2017, the situation switched, and the light curves showed a negative O'Connell effect.
  Asymmetries in the light curves seem to be a common feature in contact or near-contact binaries
  [e.g. RZ Com \citep{qianandhe}, V523 Cas \citep{zhangandzhang}, TY Boo \citep{christo},  V1104 \citep{liu}, HH UMa \citep{wang}].
    The light-curve asymmetries are usually related to changes in the surface distribution of star spots, 
    the flip-flop mechanisms  and the mass transfer between the components.

\subsection{Binary modeling}
\label{sec:mod}

The modeling procedure was, in some aspects, similar to the procedures described in \citet{fox}.
{ We used the stable build of the PHOEBE (PHysics of Eclipsing Binaries) modeling software \citep[version 0.31a]{prsa}, which is a
graphical front-end interface to the Wilson-Devinney code \citep{wilson}.}
Before running PHOEBE, the light curves were premodeled 
with Binary Maker 3.0\footnote{www.binarymaker.com} \citep{bradstreet1}, and the obtained fits in
all filter bands were input into a $B$, $V$, and $R$ simultaneous 
 modeling using PHOEBE.

The marked asymmetry and obvious variations of the light curves discussed in the previous section
strongly suggest the presence of surface spots, some degree of contact, and mass transfer
between components. Moreover, the differences in minima are moderate suggesting either semi-detached or
shallow contact configurations with the primary component filling its Roche lobe.
Therefore,  both ''overcontact'' and ''semi-detached'' modes were considered.

 We assumed a few fixed parameters during
the fitting process:  the period of the system was fixed to $P = 0.345225$ days.
The temperature of the primary star, defined as the hotter star 
producing the deeper eclipse at phase zero, was set to $T_{1}=5200$ K  based on 
spectroscopic observations discussed in Sect.~\ref{sec:spec}. 
The bolometric albedos $A_{1} = A_{2} = 0.5$ \citep{rucinski} and the
 gravity-darkening coefficients $g_{1} = g_{2} = 0.32$  \citep{lucy} appropriate for
 convective envelopes, were utilized. The logarithmic limb-darkening law is used with coefficients adopted from tables by \cite{vanhamme} 
for a solar composition star.

\begin{figure}[]
\includegraphics[width=8.5cm]{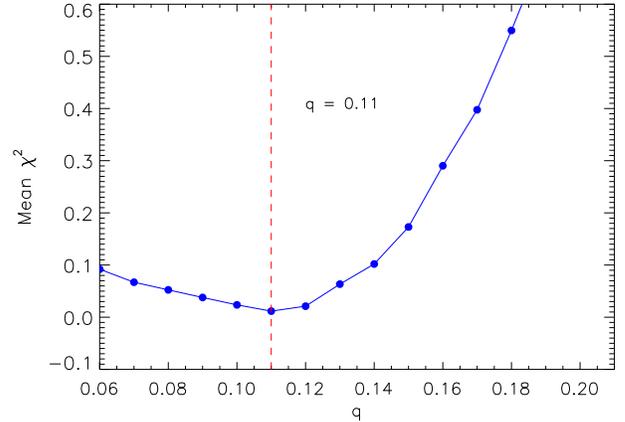}
\caption{Mean $\chi^{2}$ for unspotted model vs. a number of fixed mass ratio $q$.}
\label{fig:chi2}
\end{figure}

 We started by modeling the light curves obtained
in 2014. As it is impossible to measure component's masses in a non-eclipsing, single-lined spectroscopic binary 
like \av, a $q$-search method was applied to determine an initial mass ratio following \citet{ren}. Trial solutions of unspotted models 
with different $q$-values 
ranging from 0.05 to 0.25 were obtained. The adjustable parameters were:
the orbital inclination, $i$, the temperature of star 2, $T_{2}$;
the monochromatic luminosities of stars, $L_{1}$, and $L_{2}$; and the dimensionless potential
of each star, $\Omega_{1}$ and $\Omega_{2}$ ($\Omega_{1} = \Omega_{2}$ for  overcontact configuration).
For each mass ratio $q$ we ran  PHOEBE differential corrections in mode  ``Overcontact binary of the W~UMa type '' and then switched the 
calculations into  mode  ``Semi-detached binary, primary star fills Roche lobe''.   A set of convergent numerical solutions corresponding to
each assumed mass ratio $q$ was computed after many iterations of the adjustable parameters.
The relation between the mean $\chi^{2}$ and the assumed $q$  is described with solid circles in Figure~\ref{fig:chi2}. It can be seen 
that the resulting mean $\chi^{2}$ value of the convergent solutions reached its minimum for $q=0.11$.
We then set this initial value of $q$  and treated it as a free parameter 
along with the other adjustable parameters. 

 The  spot parameters were roughly determined with PHOEBE by adjusting the synthetic light curves to fit approximately 
the observational data. Once we had a set of approximate solutions, we ran the PHOEBE 
 differential corrections algorithm to perform the minimization 
for the input values applying the corrections to the parameter table. This was repeated several times until the lowest $\chi^{2}$ 
was obtained.  As a result, we found a best-fit binary model yielding a set of geometrical system parameters that simultaneously 
 fit the light curves in the three passbands. In the same way, we synthesized the light curves obtained in 2016 and 2017.
 
\begin{figure*}[]
\includegraphics[width=15.5cm]{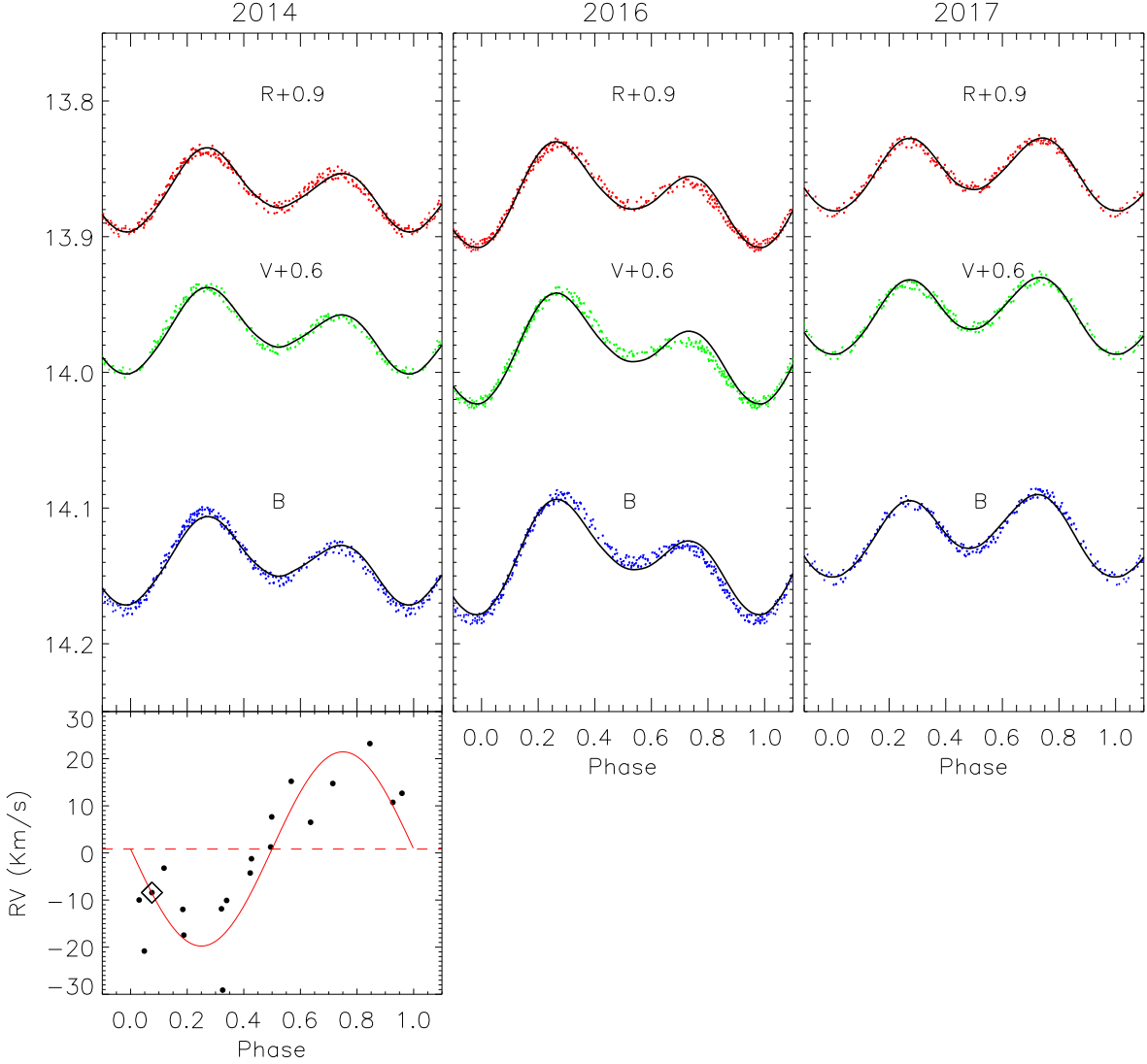}
\caption{$B$, $V$, and $R$  phase-folded light curves of \av with a period of 0.345225 days observed in 2014 (left panels), 2016 (central panels), and 2017 (right panels).
The solid lines represent the theoretical light curves of the best-model fit derived by using the PHOEBE code. Also shown are the phase-folded, radial-velocity curve with the photometric period (bottom left panel)
along with the circular orbit solution (solid lines). The rhombus indicates a LAMOST radial-velocity measurement for \av. The corresponding residuals
are shown in Figure~\ref{fig:ph3}.}
\label{fig:ph2}
\end{figure*}

\begin{figure}[]
\includegraphics[width=8.5cm]{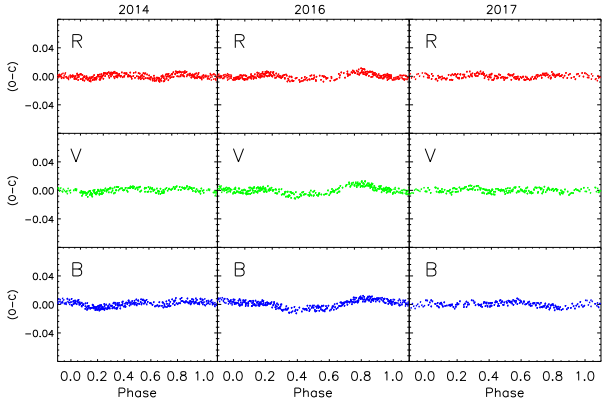}
\caption{Residuals for the best-model solutions of \av $BVR$ light curves displayed in Figure~\ref{fig:ph2}.}
\label{fig:ph3}
\end{figure}

\begin{figure}[!t]
 \includegraphics[width=8.5cm]{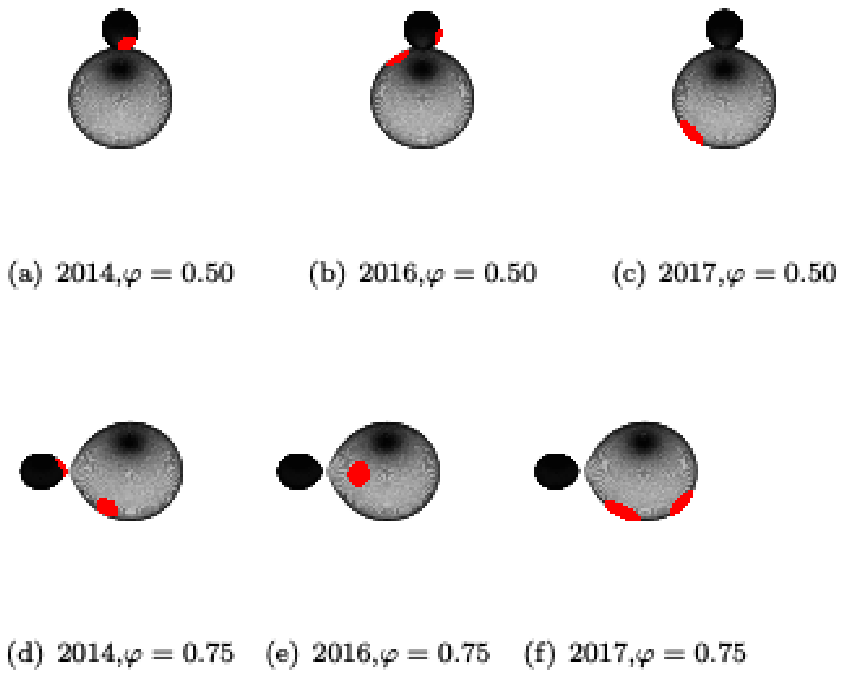}
   \caption{Geometric configurations and spots distribution of \av at phases 0.50 (top) and 0.75 (bottom).}
   \label{fig:3d}
\end{figure}

 From our analysis in overcontact mode, we concluded that all models either converged to semi-detached region or
 they were unrealistic from the evolutionary points of view, with a $q$  too small and a component temperature difference
 of $\sim$ 400 K, somewhat large for an overcontact binary. Moreover the overcontact solutions always yielded a small negative fill-out factor value. 
 Therefore, the overcontact solutions were ruled out. The parameters of the best solutions, which correspond to a semi-detached configuration, are listed in Table~\ref{tab:sol}. 
The errors for each parameter where determined with the phoebe-scripter, following
\citet{bonanos}. The theoretical light curves (solid lines) of each year
along with observations (symbols) are displayed in Figure~\ref{fig:ph2}.
The corresponding residuals are plotted in Figure~\ref{fig:ph3}.
The geometrical structures at phases 0.50 and 0.75 corresponding to the light curves solutions
 are plotted in Figure~\ref{fig:3d}.
 
As shown in Section~\ref{sec:spec},  the radial-velocity data model yields a systemic velocity $\gamma$ = 1 $\pm$ 3  Km s$^{-1}$ and 
a semi-amplitude of K$_{1}$ = 21 $\pm$ 5  Km s$^{-1}$. Combining the photometric solutions with the spectroscopic
results, the absolute parameters of mass, radius, and luminosity for each component of the system
were derived and are listed in Table~\ref{tab:abspar}.  Based on Kepler's third law, $m_{1} + m_{2} = (4 \pi^{2} /G ) (a^{3}/P^{2})$, and the
relative radius formula $r=R/a$, and luminosity $L= 4 \pi \sigma R^{2} T^{4} $, the errors of these parameters are calculated by the
error propagation method.

\begin{table*}[!t]\centering
  \setlength{\tabcolsep}{1.0\tabcolsep}
 \caption{Solution Parameters for \av. $^a$assumed.} \label{tab:sol}
  \begin{tabular}{lccc}
\hline
\hline
&&\\
Parameters  & 2014& 2016 &2017   \\

& & \\
\hline
&&\\
Mass ratio $q=m_{2}/m_{1}$         & $0.111\pm 0.05$  &$0.111 \pm 0.111$ &$0.111 \pm 0.06$ \\
$T_{1}$ (K)                        &5200$^{a}$ &    5200$^{a}$&5200$^{a}$\\
$T_{2}$ (K)                         &$4750 \pm 65$ &  $ 4700 \pm 62$ & $4709 \pm 45$ \\
$i$ (deg)                    & $38.5 \pm 0.9$ & $38.7 \pm 0.3$ &$38.3 \pm 0.5$\\
$L_{1}/(L_{1}+ L_{2})_{B}$          &$0.657 \pm 0.009$ & $0.514 \pm 0.007$ &$0.673 \pm 0.008$\\
$L_{1}/(L_{1}+ L_{2})_{V}$          &$0.799 \pm 0.007$ & $0.674 \pm 0.005$ &$0.806 \pm 0.002$\\
$L_{1}/(L_{1}+ L_{2})_{R}$           &$0.865 \pm 0.008$   & $0.747 \pm 0.009$& $0.856\pm 0.009$\\
$\Omega_{\rm in}$                   &  $1.994 \pm 0.005$ & $1.993 \pm 0.005$& $1.993 \pm 0.005$ \\
$\Omega_{1}$                         & $2.000 \pm 0.005$ & $1.999 \pm 0.007$&   $1.999 \pm 0.007$   \\
$\Omega_{2}$                          &$1.998  \pm 0.006$ & $1.998  \pm 0.009$ &  $1.998  \pm 0.009$\\
&&\\
Spot1 parameters (primary star)&&\\
Colatitude (deg)   & $90.90 \pm 0.03$  & $44.15 \pm 0.02$& $106.86 \pm 0.07$\\
Longitude  (deg)  & $62.80 \pm 0.02$  & $ 45.00 \pm 0.04$   & $60.71 \pm 0.03$  \\
Radius (deg)      &$12.50 \pm 0.07$&  $13.38 \pm 0.05$ & $21.10 \pm 0.03$\\
$T_{\rm spot} / T_{\rm surf}$        &  $0.70 \pm 0.07$  &  $0.71 \pm 0.02$ &$0.91 \pm0.08$\\
&&\\
Spot2 parameters (primary star)&&\\
Colatitude (deg)   &  &   & $100.86 \pm 0.05$\\
Longitude  (deg)  &   &   & $136.94 \pm 0.03$   \\
Radius (deg)      &    &   & $17.09 \pm 0.06$\\
 $T_{\rm spot} / T_{\rm surf}$           &  &  &$1.10\; \pm \; 0.05$\\
 &&\\
 Spot parameters (secondary star)&&\\
 Colatitude (deg)   & $90.79 \pm 0.07$ &  $90.01 \pm 0.05$ &\\
 Longitude  (deg)  & $11.00 \pm 0.06$ & $65.01 \pm 0.05$ &  \\
 Radius (deg)      &$16.20 \pm 0.07$  & $21.01 \pm 0.04$& \\
  $T_{\rm spot} / T_{\rm surf}$           & $1.10 \pm 0.08$ & $1.10 \pm 0.05$ &\\
 
&&\\
$\chi^{2}$                        &   0.23& 0.28 &0.22\\  

\hline

\end{tabular}
\label{tab:sol}
\end{table*}

\section{Discussions and Conclusions}
\label{sect:4}

We have presented follow-up photometric and spectroscopic observations of the short-period binary \av.
The photometric solutions were derived for three sets of multi-color light curves: $B$, $V$ and $R$ obtained in 2014, 2016, and 2017 years.
The results indicate that \av is a near-contact binary with both the primary and the secondary almost filling their Roche lobes, and therefore
is a marginal contact NCB system. 

\begin{figure}[]
\includegraphics[width=8.5cm]{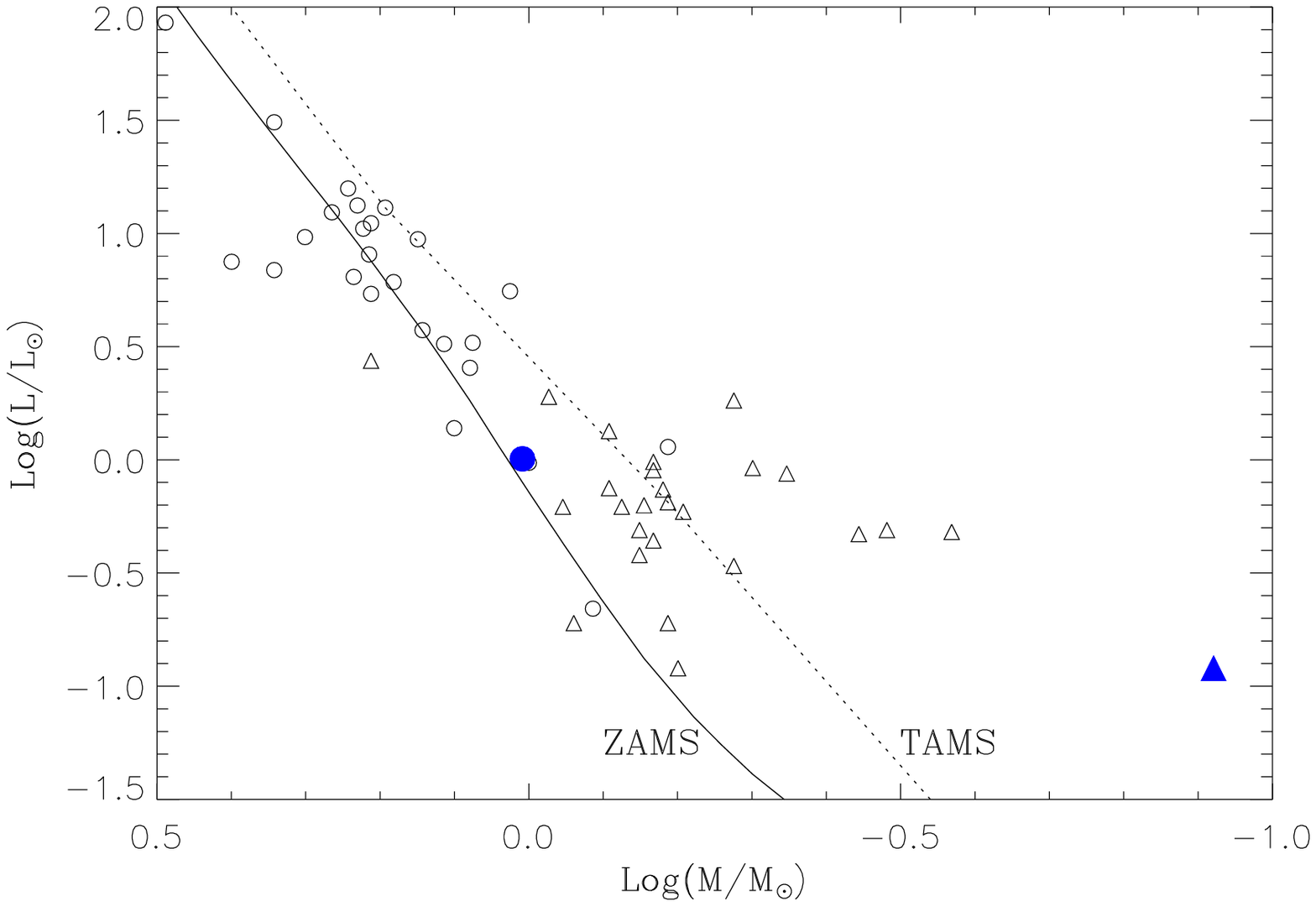}
\caption{Positions of the components of \av on the $M-L$ diagram. The filled circle denotes the primary, and the filled triangle represent
the secondary. The other symbols denote the sample of NCBs from \citet{yakut}, where  open circles and open triangles
represent the primary and secondary components. The ZAMS and TAMS are the continuous and dotted lines, respectively.}
\label{fig:evol}
\end{figure}

The spectroscopic solutions suggest that \av is a single-lined spectroscopic binary, 
 in which the primary component, partially eclipsed at phase 0.0,  is hotter than less-massive companion by $\sim$ 500 K.
 A single radial-velocity measurement reported by the LAMOST team is
in good agreement with our measurements. The photometric light curves show remarkable asymmetries and variations from one year 
to another, suggesting that the system is strongly active. The three sets of photometric solutions consider spots to account for the variations observed in the light curves.
 Combining photometric and spectroscopic solutions, we have derived the parameters listed in Table~\ref{tab:abspar}.
 The positions of both components are shown with filled symbols in the mass-luminosity diagram of Figure~\ref{fig:evol}, 
 where open circles and triangles represent, respectively, the primaries and the secondaries of NCBs from \citet{yakut}. 
 The solid and dotted lines refer, respectively, to the zero-age main sequence (ZAMS) and the terminal-age main sequence (TAMS), 
 which were constructed by the binary stars evolution (BSE) code \citep{hurley}.   It is found that the primary component of \av  is near the ZAMS line,  in agreement with our spectral classification. 
 Meanwhile, the less-massive component lies above the TAMS line.
 
A possible secular decrease in its orbital period has been found from a few times of minima over eight years time interval.
From the quadratic term of the Equation (2), it follows that the orbital period may be decreasing with a rate of 
$dP/dt = -3.87 \times 10^{-6}$ days yr$^{-1}$  or 0.3 s yr$^{-1}$.  
Such a rapid period decrease has been noticed before in others NCBs (e.g., \citealp{lohr}, or Table~5 of paper by \citealp{zhu}).  
   With the orbital period decrease, the primary component
transfer mass to the secondary one, the mass ratio increases and eventually
the NCB systems evolve into the contact phase.
Period decrease  might be caused by mass or angular momentum loss (AML) due to magnetic stellar wind (magnetic breaking) and/or
 mass transfer from the more to the less-massive component.
 
 According to the formula given by \citet{bradstreet}:

\begin{equation}
\begin{split}
 (\dot P)_{\rm AML} \sim -1.1 \times 10^{-8} q^{-1} (1 + q)^{2}  
 \times \\
 (M_{1} + M_{2})^{-5/3} k^{2} (M_{1}R_{1}^{4} + M_{2}R_{2}^{4})P^{-7/3}.
 \end{split}
\end{equation}

\noindent Using $M_{1}$, $M_{2}$, $R_{1}$ and $R_{2}$ from Table~\ref{tab:abspar}, $q=0.11$ and gyration constant $k^{2} = 0.1$,
the theoretical AML is estimated to be $(\dot P)_{\rm Theoretical} \sim -2.0\times10^{-7}$ days yr$^{-1}$.
This suggests that magnetic breaking is not the main cause of period decrease. 

The mass transfer from $M_{1}$ to $M_{2}$ or the mass loss from the system can be evaluated using the equations
given by \citet{hilditch}:

\begin{equation}
 \frac {\dot P}{P} = \frac {3 \dot M_{1}(M_{1} - M_{2})}{M_{1}M_{2}}, 
\end{equation}

\begin{equation}
 \frac {\dot P}{P} = 3{\dot M_{1}}\left[ \frac{(M_{1} + M_{2})}{M_{1}M_{2}} \frac{d^{2}}{a^{2}} - \frac{M_{2}}{M_{1}(M_{1} + M_{2})} \right ] ,
\end{equation}

\noindent for conservative and non-conservative mass loss, respectively. 
$P$ and ${\dot P}$ are the orbital period and its change rate,
$d$ is the distance from the binary centre of mass to the Lagrange point, $L_{2}$.
Calculating the necessary values of $\dot M_{1}$ to explain the observed ${\dot P}/P$,
 we obtain for \av   $\dot M_{1} \sim -4.60 \times 10^{-7}$ $M_{\odot}$ yr$^{-1}$ 
and $\dot M_{1} \sim -3.15 \times 10^{-7}$ $M_{\odot}$ yr$^{-1}$  for conservative and non-conservative mass transfer, respectively.
The timescale of the conservative mass transfer can be estimated to be approximately
$2.2\times 10^{6}$ yr. On the other hand, the thermal timescale of the massive component can be estimated
as $\tau_{th} \approx 3.0 \times 10^{7} (M/M_{\odot})^{2} (R/R_{\odot})^{-1} (L/L_{\odot})^-1$ $\sim$ $1.62 \times 10^{7}$ yr \citep{hilditch}, which
is longer than the conservative mass-transfer duration. This suggests that the primary component cannot stay in thermal equilibrium and
the mass transfer in \av is unstable. Possible mechanisms for unstable mass transfer leading to rapid period decrease in contact
or NCB systems are discussed in \cite{rasio} and \cite{jiang}.

\begin{table}
\caption{Absolute Parameters of \av.}
\begin{tabular}{lcr}
\hline
\hline
Parameter         & Primary &  Secondary         \\
                  & &            \\
 \hline
  Mass $(M_{\odot}$)  &  $1.02 \pm 0.06$ &     $0.11  \pm 0.08$\\
  Radius  $(R_{\odot}$)  &  $1.23 \pm 0.05$ & $0.45 \pm 0.05$   \\
  Luminosity $(L_{\odot}$)    & $1.01 \pm 0.06$ &$0.10 \pm 0.06$    \\
       
\hline
\hline
\end{tabular}
\label{tab:abspar}
\end{table}

An independent estimation of the distance to \av can be obtained from the empirical relation
by \citet{gettel}:

\begin{equation}
 \log D= 0.2V_{\rm max} - 0.18 \log(P) - 1.60 (J-H) + 0.56,
\end{equation}

\noindent  assuming $V_{\rm max}= 13.338$, the distance is estimated to be $\sim$ 387 pc which is in agreement with the distance 
estimated from Str\"omgren photometry by \citet{fox}, and is consistent with the fact that \av is not a cluster member  of Melotte~111 ($d \sim$ 87 pc).
 As it is known, Melotte~111 open cluster presents an apparent deficit of low mass stars, in particular {\rm K}-dwarfs (e.g. \citealp{garcia}; 
 \citealp{randich}; \citealp{terrien}).

 Our analysis indicates that \av could be in a very important evolution phase of the
thermal relaxation oscillation theory (TRO theory; e.g. \citealp{lucy}; \citealp{flannery}; \citealp{robertson}; \citealp{lucyandwilson}).
According to TRO, contact systems must undergo oscillations around the state of marginal contact as
a result of being unable to achieve thermal equilibrium. They oscillate periodically
between a contact and a near-contact phase and alternately show EW and EB light curves.
Thus, each oscillation comprises  a contact phase followed by a semi-detached phase.
When a system lies on a near-contact phase and keeps expanding, the contact configuration will
break and the system will reach the semi-detached phase with a more-massive component
filling its Roche lobe.

We note that our high-precision photometric observations were limited to a few nights every year. While the complete photometric light curve are presented,
long-time continuous observations with additional precision timings of minimum
light are still necessary to offer more information on this system.

This work has received financial support from the Universidad Nacional Aut\'onoma de M\'exico (UNAM) under grant
PAPIIT IN100918. J.N.F. acknowledges the support from the National Natural Science Foundation of China
(NSFC) through the grant 11673003 and the National Basic Research Program of China
(973 Program 2014CB845700 and 2013CB834900). Based upon observations carried out at the 
Observatorio Astron\'omico Nacional on the Sierra San Pedro M\'artir (OAN-SPM), Baja California, M\'exico, and
Xinglong Station of the National Astronomical Observatories of Chinese Academy of Science.
Special thanks are given to the technical staff and night assistants of the San Pedro M\'artir Observatory and Xinglong Station.
 L.F.M. would like to thank Profs. X.B. Zhang and J.N. Fu for their hospitality during his visit to the National Astronomical 
 Observatory of China (NAOC) and Beijing Normal University (BNU).  
 T.Q.C. and C.Q.L. would like to thank Drs. L. Fox-Machado and R. Michel for their support during a work visit
in Mexico at Insituto de Astronom\'{\i}a, Campus Ensenada.
 This research has made use of the SIMBAD database operated at CDS,
Strasbourg (France).

{\it Facilities:} 
\facility{OAN-SPM 0.84 m (MEXMAN), 2.12 m (Boller \& Chivens)}
\facility{XL 0.85 m}

\end{document}